\documentclass[traditabstract]{aa}  
\usepackage{graphicx}
\usepackage{hyperref}
\usepackage[varg]{txfonts}
\usepackage{natbib}
\bibpunct{(}{)}{;}{a}{}{,} 

\begin{document}

\title{Crazy heart: kinematics of the ``star pile'' in Abell
  545\thanks{Based on observations taken at the European Southern
    Observatory, Cerro Paranal, Chile, under programme ID
    080.B-0529. Also based on observations obtained at the Gemini
    Observatory, which is operated by the Association of Universities
    for Research in Astronomy, Inc., under a cooperative agreement
    with the NSF on behalf of the Gemini partnership: the National
    Science Foundation (United States), the Sscience and Technology
    Facilities Council (United Kingdom), the National Research Council
    (Canada), CONICYT (Chile), the Australian Research Council
    (Australia), Minist\'erio da Ci\^encia e Tecnologia (Brazil) and
    SECYT (Argentina); and on observations obtained with
    MegaPrime/MegaCam, a joint project of CFHT and CEA/DAPNIA, at the
    Canada-France-Hawaii Telescope (CFHT) which is operated by the
    National Research Council (NRC) of Canada, the Institut National
    des Science de l'Univers of the Centre National de la Recherche
    Scientifique (CNRS) of France, and the University of Hawaii.}}

\author{
R.~Salinas\inst{1,2}
\and 
T.~Richtler\inst{1}
\and  
M.~J.~West\inst{2}
\and
A.~J.~Romanowsky\inst{3}
\and
E.~Lloyd-Davies\inst{4}
\and 
Y.~Schuberth\inst{5}
}

\offprints{Ricardo Salinas}

\institute{
Departamento de Astronom\'{\i}a, 
Universidad de Concepci\'on, 
Concepci\'on, 
Chile; [rsalinas,tom]@astro-udec.cl
\and
European Southern Observatory, 
Alonso de C\'ordova 3107, 
Santiago, 
Chile; mwest@eso.org
\and
UCO/Lick Observatory, 
University of California, 
Santa Cruz, CA 95064, 
USA; romanow@ucolick.org
\and
Astronomy Centre, 
University of Sussex, 
Falmer, Brighton, BN1 9QH, 
UK; E.Lloyd-Davies@sussex.ac.uk
\and
Argelander Institut f\"ur Astronomie, 
Auf dem H\"ugel 71, 
53121 Bonn, 
Germany; yschuber@astro.uni-bonn.de
     }
   \date{Received ; accepted }

  \abstract{
   
    We study the structure and internal kinematics of the ``star
    pile'' in Abell 545 -- a low surface brightness structure lying in
    the center of the cluster.
     We have obtained deep long-slit spectroscopy of the star pile
     using VLT/FORS2 and Gemini/GMOS, which is analyzed in conjunction
     with deep multiband CFHT/MEGACAM imaging.
     As presented in a previous study the star pile has a flat
     luminosity profile and its color is consistent with the outer
     parts of elliptical galaxies. Its velocity map is irregular, with
     parts being seemingly associated with an embedded nucleus, and
     others which have significant velocity offsets to the cluster
     systemic velocity with no clear kinematical connection to any of
     the surrounding galaxies. This would make the star pile a
     dynamically defined stellar intra-cluster component.
     The complicated pattern in velocity and velocity dispersions
     casts doubts on the adequacy of using the whole star pile as a
     dynamical test for the innermost dark matter profile of the
     cluster. This status is fulfilled only by the nucleus and its
     nearest surroundings which lie at the center of the cluster
     velocity distribution.

}
\titlerunning{Kinematics of the star pile in Abell 545}

   \maketitle
%

\section{Introduction}
\label{sec:intro}
The central regions of massive galaxy clusters are fascinating places
to study.  Spiral galaxies are not allowed to exist in these
high-density regions \citep{dressler80,whitmore93}. Processes such as
mergers, cooling flows and tidal stripping work together to often form
large, extended cD galaxies
\citep[e.g.][]{ostriker75,cowie77,richstone76}. Not only the density
of galaxies, i.e. baryonic dark matter, but also the density of the
still mysterious dark matter is highest in the centers of massive
clusters.  A cluster-wide cosmological dark halo \citep{maccio08} of
the NFW type with a virial mass of $10^{15} M_\odot$ and a virial
radius of 2 Mpc encloses in its central 10 kpc a mass of
$2\times10^{11} M_\odot$ and in its inner 1 kpc a mass of
$2\times10^{9} M_\odot$. These values are equal to or even higher than
the baryonic mass of a normal central galaxy, so one should expect
dynamical situations which are very different from those in lower
density regions.

An example is Abell 545. This galaxy cluster is one of the richest and
most massive clusters in the Abell catalogue. Accordingly it has
  a high global velocity dispersion of $\sim$\,1150 km s$^{-1}$ (Salinas
  et al. in preparation). The central region is populated by a dense
configuration of medium-bright early-type galaxies (about 25 within
projected 100 kpc) while no cD galaxy is present. Instead one finds an
extended, elongated structure of low surface brightness with a major
axis of about 60 kpc, dubbed the ``star pile'' \citep{struble88}.
Embedded in the star pile are at least three dwarf-like galaxies, the
middle one being projected onto the geometrical center of the star
pile (Fig. \ref{fig:longslits}).  The star pile's stellar nature and
that it indeed belongs to Abell 545, has been shown by
\citet[][hereafter S+07]{salinas07}, who also advertised the star pile
as a dynamical probe for the inner dark matter profile shape. This is
interesting because there is evidence that the slopes of the central
dark matter profiles in some galaxy clusters are shallower than those
found in simulations \citep[][Richtler et
al. 2010]{kelson02,sand04,sand08}.  \defcitealias{salinas07}{S+07}

Since the baryonic density of the star pile is low, one expects the
entire inner region of Abell 545 to be dark matter dominated and the
dynamics of the star pile could provide important insight into the
problem of whether the inner region of the dark matter halo has a cusp
or a core, testing the prediction of cuspy profiles of cosmological
dark matter simulations \citep[e.g.][]{navarro96}. Such a faint
structure may even avoid the problems that arise due to the dynamic
interactions between baryons and dark matter when a central galaxy is
present \citep{elzant04,gnedin04,delpopolo09}.  However, its usage as
dynamical tracer relies on the assumption that the star pile is at
rest at the cluster center and that its velocity dispersion is the
result of a dynamical system in equilibrium.

Until now we had no knowledge of the kinematical situation in the core
region of Abell 545.  The intention of this contribution is therefore
to present kinematic data for the star pile and the innermost galaxies
(velocities and velocity dispersions).  
For Abell 545, we adopt the redshift $z$\,=\,0.1585 (Salinas et al. in
preparation), which corresponds (in a standard flat universe) to an
angular distance of 565 Mpc and a scale of 2.74 kpc/$\arcsec$.

\section{Observations and data reduction}
\subsection{CFHT/MegaCam imaging}
\label{sec:megacam}

In \citetalias{salinas07} we presented photometry of the star pile
using $VRI$ archival images obtained with FORS1. Here we revisit the
star pile making use of deeper images retrieved from the
MegaPrime/MegaCam archive (PI: G. Morrison). MegaCam is an optical
camera installed at the Canada-France-Hawaii Telescope at Mauna
Kea. MegaCam images have a field-of-view of $\sim$1 deg$^2$ and a
0.187 arcsec/pixel scale. The image set is composed of $ u^*g'r'i'z'$
images. The original images were obtained on several nights during
October 2005.  Details on exposure times and seeing quality can be
seen
in Table~\ref{table:megacam} of the on-line Appendix. The archive
images have been pre-processed (bias subtracted, flat fielded and
background subtracted) by the MegaCam pipeline (known as Elixir) and
stacked by the MegaCam image stacking pipeline, MegaPipe. 

Photometry of the cluster galaxies in the five Megacam bands was
performed using SExtractor \citep{bertin96}. The full cluster
photometry will be presented in a forthcoming publication focusing on
the cluster dynamics. Here we present results for the innermost
galaxies surrounding the star pile.

\onltab{1}{
  \begin{table*}
   \caption{Abell 545 MegaCam observations}
   \label{table:megacam}      
   \centering 
   \begin{tabular}{l c c c c c}        
   \hline\hline                 
    &$u^*$ & $g'$ & $r'$ & $i'$ &$z'$ \\    
   \hline                        
    Exp. time ($s$)& $6\times720$,$1\times72$ & $5\times720$ & $5\times720$ & $5
\times720$ & $6\times720$\\   
    $FWHM$ (in \arcsec)&   1.27 & 1.18 & 0.95 & 1.0 & 0.68\\
    Magnitude limit (5$\sigma$)&26.5 &27.1&26.3&25.5&24.8\\
   \hline                                  
   \end{tabular}
   \end{table*}
}

\subsection{Gemini/GMOS long-slit spectroscopy}
\label{sec:gemini}

Long-slit observations with Gemini South/GMOS (Cerro Pach\'on, Chile)
of the star pile in the center of Abell 545 were carried out during
October and November 2007 (program ID GS-2007B-Q-9). Exposure time was
12$\times$2180 secs with PA\,=\,209.3 degrees East from North, roughly
the star pile's major axis, and 12$\times$2180 secs with PA\,=\,151.6
degrees East from North, an intermediate axis of the star pile (see
Fig. \ref{fig:longslits}). We used the B600+\_G5323 grating and a slit
width of 1\arcsec\ which gives a resolution of $\sim$6.2\,\AA\ FWHM.

The spectroscopic reductions were carried out with the aid of the
Gemini/IRAF package (version 1.9.1). Bias subtraction, flat fielding
and wavelength calibration were performed in the standard way. Typical
residuals from the wavelength calibration were $\sim$0.07\,\AA.
Subsequently, the zero-point of the dispersion solution was checked by
measuring the bright sky line at 5577.34\,\AA\ and corrected when
necessary. These corrections were always less than 0.1\,\AA.  The
spectra for each position angle were combined applying a median filter
with the IRAF task \verb+lscombine+. These median 2D spectra were sky
subtracted using the \verb+background+ task inside IRAF.

\subsection{FORS2 long-slit spectroscopy}
\label{sec:fors2}
\begin{figure}[tp]
   \centering
   \includegraphics[width=0.49\textwidth]{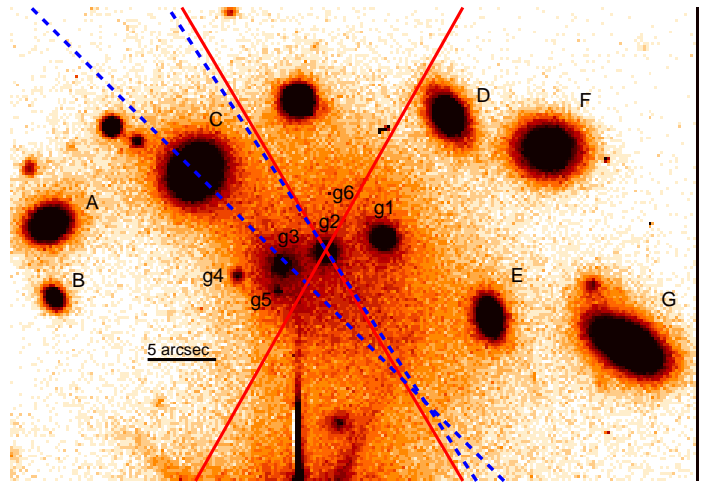}
   \caption[Star pile long-slit positioning]{A FORS1 V image excerpt
     showing the star pile and the longslit positioning. Red lines
     indicate the position of the GMOS longslits, while the blue
     dashed lines indicate the position of the FORS2 longslits. Faint
     sources, seemingly embedded in the star pile have been labeled
     \emph{g1-g6}. Surrounding galaxies are labeled with letters A to
     G. The size of the image is 50\arcsec\,$\times$\,35\arcsec or
     137\,$\times$\,96 kpc at the cluster's distance. North is up and East
     is to the left.}
   \label{fig:longslits}
   \end{figure}

Further long-slit observations were obtained with the Very Large
Telescope (VLT) at Cerro Paranal, Chile, using the FORS2/MXU
instrument in service mode (program ID 080.B-0529) between November
2007 and March 2008. Long-slit observations were taken with two
different position angles: PA\,=\,33.2 degrees East from North, roughly
coincident with the apparent photometric major axis of the star pile,
and PA\,=\,43.0 degrees East from North, an intermediate axis passing
through the eastern apparent nucleus of the star pile and the center
of one of the neighbor giant elliptical galaxies that surround the
star pile (labeled as C in Fig. \ref{fig:longslits}). The chosen grism was
GRIS\_1200R+93 with a slit width of 1.31\arcsec\, which gives a
resolution of $\sim$4\,\AA\, FWHM in the 5750--7200\,AA\, wavelength
range.  Both slit positions had an exposure time of 18$\times$1333
seconds each, i.e. $\sim$6 hours and 40 minutes.

FORS2 long-slit observations were bias subtracted, flat fielded and
wavelength calibrated with the ESO FORS pipeline. Since a large number
of exposures were to be combined, no attempt at eliminating cosmic
rays on individual images was made. Since the star pile was located on
the ``master'' chip that is expected to give a better performance,
only this chip was reduced and further analyzed. Exposures for each PA
were median combined with IRAF/\verb+imcombine+.
Sky subtraction done by the ESO FORS pipeline was unsatisfactory so it
was performed with the IRAF task \verb+background+ by applying a
median filter to a clean portion of the sky approximately 50\arcsec\
NE of the star pile for the PA\,=\,33.2$\degr$ longslit and 80\arcsec\
NE for PA\,=\,43$\degr$.

The extraction of the FORS2 and GMOS long-slit spectra was done with a
custom IDL task. Since the signal from the star pile is weak, it was
possible to extract only between one to three radial bins at each side
of the ``nucleus'' inside each slit. Both direct sums and median were
used for the spatial binning. Since the surface brightness within the
radial bins is not constant, but significantly decreasing due to the
large bin size (see Fig. \ref{fig:profiles}), the median was much
noisier and unrepresentative of the star pile as a whole, compared
with the direct sum, which was therefore chosen for the analysis.

   \subsection{GMOS and FORS2/MXU multi-object spectroscopy}
\label{sec:mos}
Additional multi-object spectroscopy of cluster galaxies was taken
using GMOS and FORS2.  Details of these observations will be presented
in a forthcoming publication describing the dynamics of the
cluster. Here we present only velocities for the ``nuclei'' of the
star pile as well as its nearest neighbors.

\section{Photometric properties of the star pile}
\label{sec:photometry}
\citetalias{salinas07} presented $VRI$ photometry of the star pile
from archival FORS1 images. The photometry was measured using
IRAF/\verb+ellipse+, using the galaxy \textit{g2} as the center. The
main photometric results from \citetalias{salinas07} are: a) the
luminosity profile of the star pile is
flatter than a typical cD halo, b) there
is no evidence for a color gradient inside the star pile; the colors
are consistent with an old and metal rich population; and c) the
integrated luminosity, including the contribution from \textit{g2},
resembles that of a giant elliptical.

\begin{figure}[t!]
     \centering
     \includegraphics[width=0.5\textwidth]{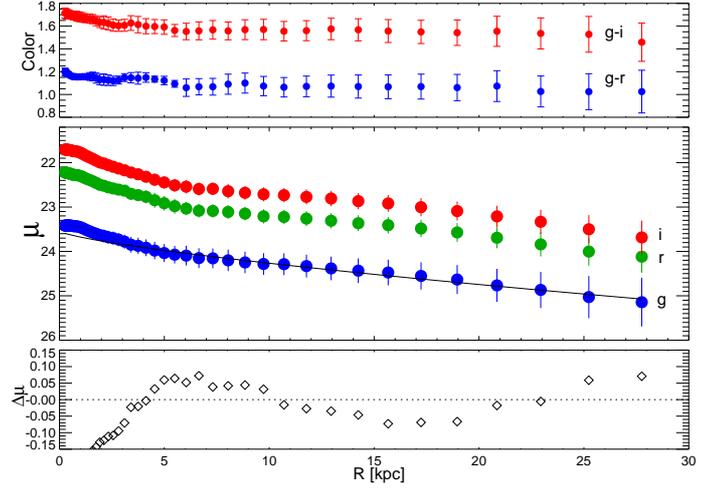}
     \caption[Star pile surface photometry]{Azimuthally averaged
       $g'r'i'$ surface brightness profile of the star pile from the
       MEGACAM images. The top panel shows $g'-r'$ and $g'-i'$ color
       profiles, while the lower panel shows the residuals from the
       S\'ersic fit to the $g'$ photometry.}
     \label{fig:profiles}
   \end{figure}

The new spectroscopic data however suggest that the star pile is
associated with the galaxy \emph{g3} rather than with \emph{g2} (see
Sect. \ref{sec:nuclei_kin}). We re-do the star pile photometry, this
time centered on \textit{g3}, using the MEGACAM images, which although
obtained under worse seeing conditions, are deeper than the FORS1
images.  As with the FORS1 images, the photometry was done with
IRAF/\verb+ellipse+, masking neighbor galaxies. For the innermost
radii, center, ellipticity and position angle were let free to vary
until a radius of 1.5\arcsec, where the fit became unstable. Beyond
that radius, ellipticity and position angle were kept fixed at
$\epsilon$\,=\,0.15 and PA\,=\,33$\degr$, and only the center of the
ellipses was allowed to vary. $g'r'i'$ photometry can be seen on
Fig. \ref{fig:profiles}. The main difference with the FORS1 photometry
presented in \citetalias{salinas07} is that the transition in the
light profile from the \emph{g3} nucleus to the star pile is smoother,
indicating perhaps a physical connection, that is not the case of
\emph{g2}.

The main findings from \citetalias{salinas07} are still valid; the
profile presents a mild color gradient in the inner 5 kpc (which can
be associated with the nucleus), that can be partially attributed to
the slight difference between the image quality in the different
photometric bands. Beyond 5 kpc the color profiles are flat within the
errors, with values $(g'-r')_0$\,=\,0.94 and $(g'-i')_0$\,=\,1.33,
where we have corrected for the Galactic absorption
\citep{schlegel98}. Subtraction of the light model does not reveal any
hidden substructure, for example, gravitational arcs.

   We fit a S\'ersic profile \citep{sersic68,caon93} to the $g'$
   photometry, which is less affected by bleeding from the bright star
   located 20\arcsec\ South of the star pile (see
   Fig. \ref{fig:longslits}), outside the inner arcsecond which is
   affected by seeing. The photometry, excluding the innermost
   arcsecond (2.74 kpc at the cluster's redshift) is well reproduced
   by S\'ersic parameters $\mu_e$\,=\,26.15, $R_e$\,=\,57.3 kpc and
   $n$\,=\,1.36. The low value of $n$ indicates that the profile is
   much flatter than the profile of a typical elliptical galaxy. The
   S\'ersic fit and its residuals can be seen in
   Fig. \ref{fig:profiles}.

   We further model the star pile using a S\'ersic profile, in order
   to address its luminosity and mass profiles. First, the
   mag/arcsec$^2$ profile is transformed to a luminosity profile using
   the distance modulus of 39.39, Galactic absorption $A_{g}$\,=\,0.6
   \citep{schlegel98} and $M_{\odot,g}$\,=\,5.12 \citep{blanton07}. We
   also apply the correction for cosmological dimming, $(1+z)^4$. This
   $L_{\odot}$ pc$^{-2}$ profile is re-fit with a S\'ersic profile of
   the form,
\begin{equation} 
  I(R)=I_0 \mathrm{exp}\left[-\left(\frac{R}{a_s}\right)^{\frac{1}{m}}\right],
\end{equation}
which does not have an exact analytical deprojection, but for which a
good analytical approximation exists
\citep{prugniel97}, 
\begin{eqnarray*}
  j(r)&=&j_1\tilde{j}(r/a_s),\label{eq:sersic_dep}\\
  \tilde{j}(x)&\simeq& x^{-p}\mathrm{exp}(-x^{1/m}),\\
  j_1&=&\left\lbrace\frac{\Gamma(2m)}{\Gamma[(3-p)m]}\right\rbrace\frac{I_0}{2a_s
  },\\
 p&\simeq& 1 - 0.6097/m + 0.05463/m^2,
\end{eqnarray*}
where the latter equation comes from \citet{lima99} and $\Gamma(x)$ is
the Gamma function.
   
Integrating this luminosity density profile, $j(r)$, it is possible to
derive an enclosed luminosity. For the S\'ersic profile deprojection
as given in Eq. \ref{eq:sersic_dep}, the enclosed luminosity has again
an analytical expression given by \citet{lima99} 
\begin{eqnarray}
  L_s(r)&=&L_{\mathrm{tot}}\tilde{L}_s(r/a_s),\label{eq:sersic_lum}\\
\tilde{L}_s(x)&=&\frac{\gamma[(3-p)m,x^{1/m}]}{\Gamma[(3-p)m]},
\end{eqnarray}
where now $\gamma$ is the incomplete Gamma function
and 
\begin{equation} 
  L_{\mathrm{tot}}=2\pi m\Gamma(2m)I_0a_s^2 
\end{equation}
is the total luminosity of the profile.

\begin{figure}[tp!]
  \centering
  \includegraphics[width=0.5\textwidth]{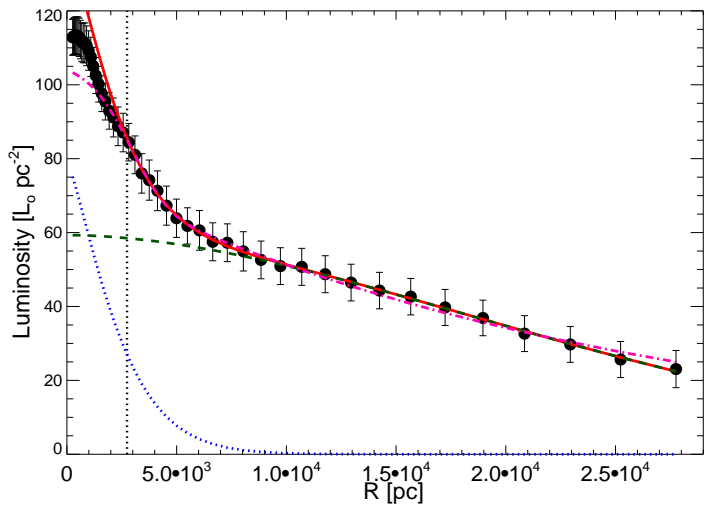}
  \caption{Star pile $g'$ luminosity profile. The blue dotted line is the
    S\'ersic fit to the nucleus, while the green dashed line is a fit
    to the extended star pile. The red solid line is the sum of both. The pink
    dot-dashed line is the S\'ersic+exponential model. The radius is the
    semi-major axis of the fitted ellipses. The vertical dotted line
    delimits the inner 1\arcsec\, that was excluded from the fit.}
  \label{fig:sp_luminosity}
\end{figure}

The light profile shows a break in the slope around 5 kpc (see
Fig. \ref{fig:sp_luminosity}). This could be interpreted as the place
where the star pile light starts to dominate over the nucleus if they
happen to be different entities.
Assuming both \emph{g3} and the star pile form distinct components, we
separate them by modeling the luminosity profile as a sum of two
S\'ersic profiles and also as a S\'ersic plus exponential profile,
which has been found to give a good fit to cD envelopes
\citep{seigar07}. Fig. \ref{fig:sp_luminosity} shows both models (red
and pink lines) as well as the components of the double S\'ersic
profile.
The fit has avoided the inner
1\arcsec\, which is affected by seeing. We compute the goodness of fit
using the rms scatter following \citet{seigar07},
\begin{equation} 
\Delta=\sqrt{\frac{\sum_i^N\delta_i^2}{N}},
\end{equation}
where $N$ is the number of points and the $\delta_i$ are the
difference in magnitudes between the model and the data.  The fit
parameters and the goodness of fit of both models can be seen in Table
\ref{table:sp_fits}. The star pile is better described with the
S\'ersic profile rather than an exponential. This reinforces the fact
that the star pile is something different than a cD envelope.

 \begin{table}
   \caption{Structural parameters of the S\'ersic+S\'ersic and S\'ersic+exponential models of the surface brightness profile of the star pile.}
   \label{table:sp_fits}      
   \centering 
   \begin{tabular}{l|ccc}        
     \hline\hline                 
     S\'ersic + S\'ersic& $I_0$ & $a_s$ & $m$\\
     $\Delta=0.40$ &[$L_\odot$ pc$^{-2}]$ &[kpc] & \\
     \hline
     inner&79.19&2.588&0.78\\
     outer&58.32&28.178&0.55\\
     \hline
     \hline
     S\'ersic + exp& $I_0$ & $a_s$ & $m$\\
     $\Delta=0.74$ &[$L_{\odot}$ pc$^{-2}]$ &[kpc] & \\
     \hline
     inner&27.27&3.46&0.35\\
     outer&76.86&24.706&1\\
     \hline                                  
   \end{tabular}
   \end{table}

   Using Eq. \ref{eq:sersic_lum} we calculate the total luminosity of
   both profiles, $L_{\mathrm{g3}}$\,=\,2.31$\times$10$^9$ L$_{\odot}$
   and $L_{\mathrm{sp}}$\,=\,5.56$\times$10$^{10}$ L$_{\odot}$, where the
   luminosity of the star pile is calculated within the inner 28 kpc,
   where the photometry has been measured. This luminosity transforms
   into an absolute magnitude of $M_{g'}\!=\!-21.75$ or $M_V=-22.35$
   using filter transformations from \citet{blanton07}, that is,
   almost as luminous as the brightest galaxies in the cluster, and in
   good agreement with the value given by \citetalias{salinas07},
   $M_V\!\sim\!-22.5$, where the measurement included the galaxy
   \emph{g2}.

   Taking the colors $(g'-i')_0$\,=\,1.33 and $(g'-r')_0$\,=\,0.94,
   together with color transformations from \citet{bell03}, we
   estimate $M/L_{g'}$\,=\,7.7 for the star pile. This means that the star
   pile has a stellar mass of $4.3\times10^{11}$ $M_{\odot}$ within 30
   kpc. Adding \textit{g3}, the baryonic mass slightly increases to
   $4.5\times10^{11}$ M$_{\odot}$.

\section{Kinematics of the star pile and its ``nuclei''}
   \subsection{Measuring the star pile kinematics}
   \label{sec:sp_kinematics1}
   With the spectra presented in \citetalias{salinas07} it was
   possible to confirm the star pile as a cluster feature, but without
   the sufficient $S/N$ in order to measure a velocity dispersion. The
   heliocentric radial velocity of the star pile was found to be
   47100$\pm$60 kms$^{-1}$, the uncertainty resulting from the
   measurements of 16 well identifiable absorption lines. No
   significant traces of velocity differences were found within the
   star pile.

   With this result in hand, the new spectroscopic observations were
   planned in order to obtain a velocity dispersion measurement of the
   star pile. The velocity dispersion of the star pile from the GMOS
   and FORS2 long-slit observations was determined with the penalized
   Pixel Fitting code \citep[pPXF,][]{cappellari04}. The program
   selects from a series of templates the best templates that, when
   convolved with an appropriate line-of-sight velocity distribution,
   best reproduce the science spectrum. Even though the program allows
   the extraction of higher order moments of the velocity
   distribution, this option was suppressed due to the modest $S/N$ of
   the star pile spectra. As templates we selected the MILES stellar
   library \citep{sanchez06} that contains 985 stars comprising a wide
   range of spectral types and metallicities.

   After the extraction of the star pile spectra as described in
   \ref{sec:fors2}, and before entering the spectra to pPXF, the
   spectra were de-redshifted using the IRAF task \verb+dopcor+ and
   using the cluster redshift, $z$\,=\,0.15855 (Salinas et al. in
   preparation). Afterwards, the spectra were cut in the
   4900--5400\,\AA\, range in the case of the FORS2 observations, and
   in the 3900--5000\,\AA\, range for the GMOS spectra, both measured
   in the cluster's rest frame. The blue edge of the FORS spectra was
   determined by the grism and the red edge was selected to avoid
   bright sky lines. In the case of the GMOS spectra, the blue edge
   was taken to include the very noticeable Ca H-K absorption lines.
   Errors in the measured quantities are obtained by adding noise to
   100 Monte Carlo realizations of the extracted spectra and running
   pPXF over them. The dispersion of the new measurements is quoted as
   the error in the quantities measured in the original spectrum. 
     Possible systematics in the pPXF measurements introduced during
     the sky subtraction procedure were explored by constructing sky
     spectra from both sides of the star pile long-slits. We added/subtracted 
      5\% of the signal to/from these sky spectra before
     subtracting from the star pile spectra, which then passed 
     pPXF using the  procedure outlined before. We find that
     velocity dispersions changed by at most 15 km s$^{-1}$, while
     changes in the mean velocities were never larger than 10 km
     s$^{-1}$ within the uncertainties given by the Monte Carlo
     procedure.

     Galaxy velocities in the Abell 545 field have been measured using
     IRAF/\verb+fxcor+. For consistency, velocities coming from the
     different star pile spectra were also measured with this task. To
     measure a velocity, the high $S/N$ that is required for a
     velocity dispersion measurement is not necessary, so instead of
     using the large spatial binning applied for the dispersion
     calculations, the star pile longslit spectra were spatially
     rebinned calculating the median of three adjacent pixels. The
     same wavelength range used for the velocity dispersion
     measurements was used for these velocity measurements. As
     templates for \verb+fxcor+ we selected 20 old, metal rich single
     stellar population models from \citet{vazdekis10}. The velocities
     and errors given in Fig. \ref{fig:ls_velocities} and in Table
     \ref{table:nuclei} were taken as the mean results from the five
     templates which gave the highest cross-correlation peak. These
     have ages between 7--14 Gyrs and metallicities
     $[M/H]$=0.0--0.2. The zero-point of the velocity calibration
     between the galaxies and the longslits was determined by the
     comparison of the measured velocities of galaxies C and \emph{g3}
     (see Fig. \ref{fig:center}) which they had in common. The
     velocities from both spectroscopic methods agree within the
     errors.

  \begin{figure}[t!]
  \centering
      \includegraphics[scale=0.48]{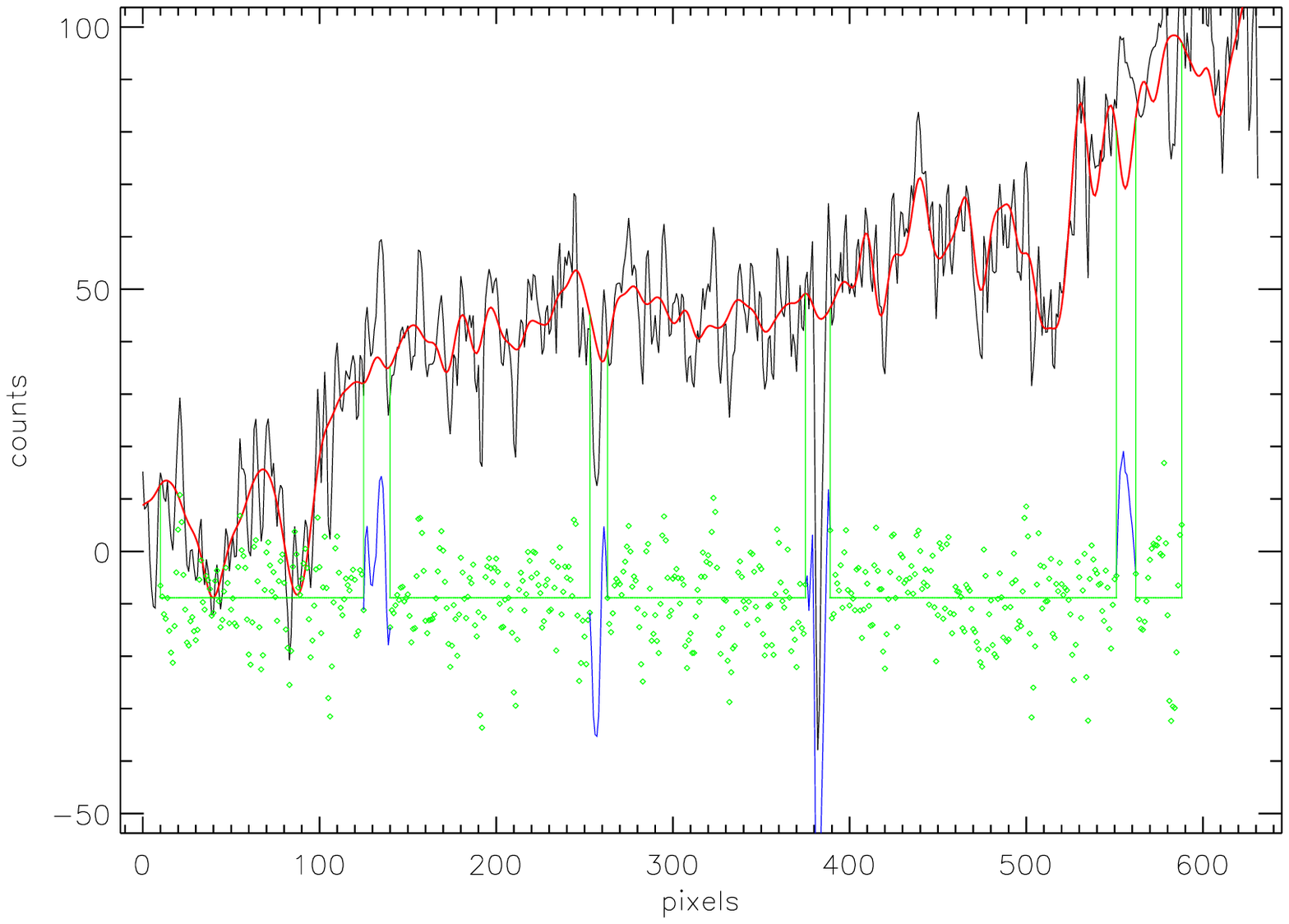}
     \includegraphics[scale=0.48]{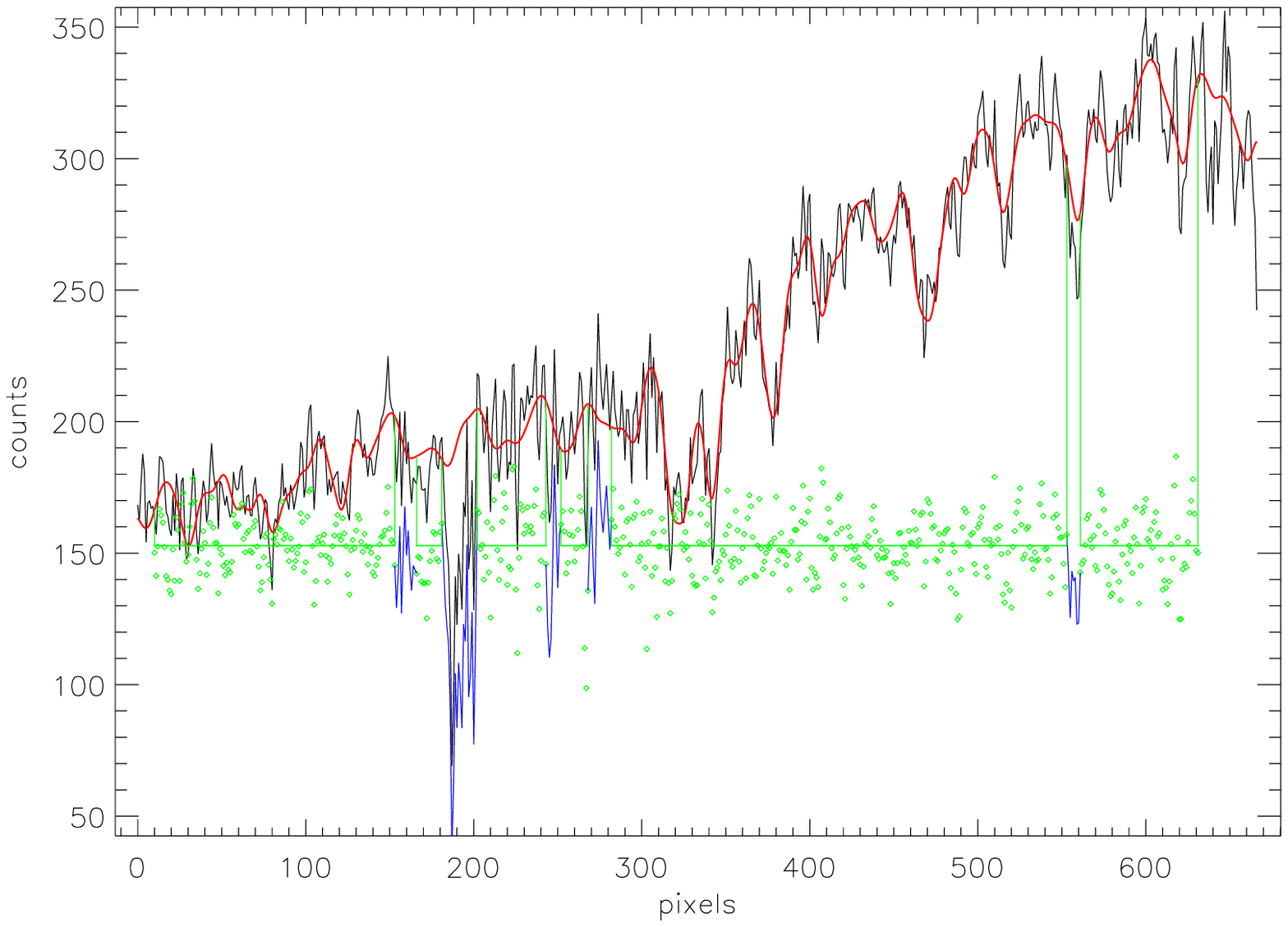}
     \caption{Examples of spectra and pPXF fits. {\bf Top panel}: the
       star pile spectrum in the 9--21 kpc range coming from GMOS
       intermediate axis longslit. {\bf Bottom panel:} \textit{g3} nucleus
       spectrum from the FORS2 data. In both panels the original
       spectrum is drawn in black, while the fit is in red. Green
       points indicate the fit residuals, and blue portions indicate
       ranges that were excluded from the fit due to their large
       residuals from ill-subtracted sky lines or other artifacts. The
       wavelength range is 3900--4400\AA\ and 4900--5400\AA\ for the
       top and bottom panels, respectively, in the rest
       frame.}\label{fig:ppxf}
   \end{figure}

   \subsection{The star pile kinematics: results}
   \label{sec:sp_kinematics2}
   The results of the pPXF fitting for the different long slits and
   the velocity determination with \verb+fxcor+ can be seen in Table
   \ref{table:vdps}. A visual representation of this table, together
   with velocities and velocity dispersions, when available, for
   neighboring galaxies, is given  in
   Fig. \ref{fig:center}. Velocity measurements with the low spatial
   binning can be seen in Fig. \ref{fig:ls_velocities}.

   \subsubsection{The star pile velocity map}
   \label{sec:sp_velocity_map}
  
 \begin{figure}[t!]
   \centering
   \includegraphics[width=0.5\textwidth]{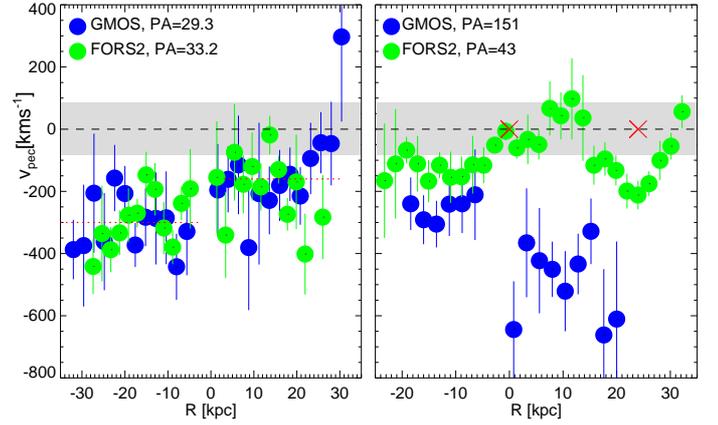}
   \caption{Velocity measurements from FORS2 (blue symbols) and GMOS
     (green symbols) with respect to the cluster velocity. The shaded
     grey area indicates the 1-$\sigma$ confidence level for the
     cluster mean redshift. The zero point of the radii is defined by
     the position of the nuclei over which the long slit is
     placed. {\bf Left panel}: major axis velocity profile, the red
     dotted lines indicate the mean velocities North and South of the
     central galaxy. {\bf Right panel}: the two intermediate axis
     velocity profiles. The red crosses in the center mark the
     positions of the galaxies \textit{g3} and ``C'' that the FORS2
     longslit crosses.}
   \label{fig:ls_velocities}
   \end{figure}

   The first, and most surprising
   result  is the fact that the star pile, or at least a substantial
   part of it, does not seem to be at rest at the bottom of the
   potential well as expected. All the long slit mean velocity
   measurements point to a velocity difference of $\sim$100--500
   km s$^{-1}$\, less than the cluster systemic velocity.
   The major axis velocity measured from the GMOS and FORS2 longslits
   agree within the errors (Fig. \ref{fig:ls_velocities}, left
   panel). We note that a significant difference exists between the
   sections of the star pile located North (positive radii in left
   panel of Fig. \ref{fig:ls_velocities}) and South (negative radii)
   of the galaxy \textit{g2} that the major axis longslit
   crosses. While the South part has a mean velocity difference with
   the cluster velocity of $-300$ km s$^{-1}$, the difference of the
   North section is $-160$ km s$^{-1}$. Since the uncertainty of the
   cluster mean velocity is $\pm$ 84 km s$^{-1}$ (Salinas et al. in
   prep), it is possible to consider some parts of the star pile as at
   rest compared the cluster. This specially applies to the zones
   nearest to \textit{g3} (see Fig. \ref{fig:ls_velocities}, right
   panel).

   The GMOS intermediate axis (blue points in the right panel of
   Fig. \ref{fig:ls_velocities}) gives the most extreme velocity
   differences. In the North-West part of the star pile, the mean
   velocity reaches a $\sim$450 kms$^{-1}$\, difference with the
   cluster velocity. A couple of measurements reach even larger
   velocities. The one at $R\sim0$ kpc, with
   $\varv\sim-600$kms$^{-1}$, is possibly contaminated by the spectrum
   of the high velocity galaxy \emph{g2}. The measurements at
   $R\sim20$ kpc are more difficult to explain. Even though the nearby
   galaxy (labeled ``D'' in Figs. \ref{fig:longslits} and
   \ref{fig:center}) has a similar rest frame velocity, $\varv=-437
   \pm 42$ kms$^{-1}$, it seems too distant to be contaminating this
   section of the star pile. It is also interesting to mention that
   this longslit crosses a very faint source, which we have labeled
   \textit{g6}, that is only visible in the FORS1 $V$ image (which has
   the best seeing in all our imaging dataset).
   It is not possible to say if this is an independent source (nucleus
   or background galaxy) or some substructure within the star
   pile. However, the source is unlikely to influence the velocity
   dispersion in the NW part of the star pile due to its low surface
   brightness.

   \subsubsection{The star pile velocity dispersion}
    \label{sec:sp_dispersion}

    \begin{figure}[t!]
      \centering
      \includegraphics[width=0.49\textwidth]{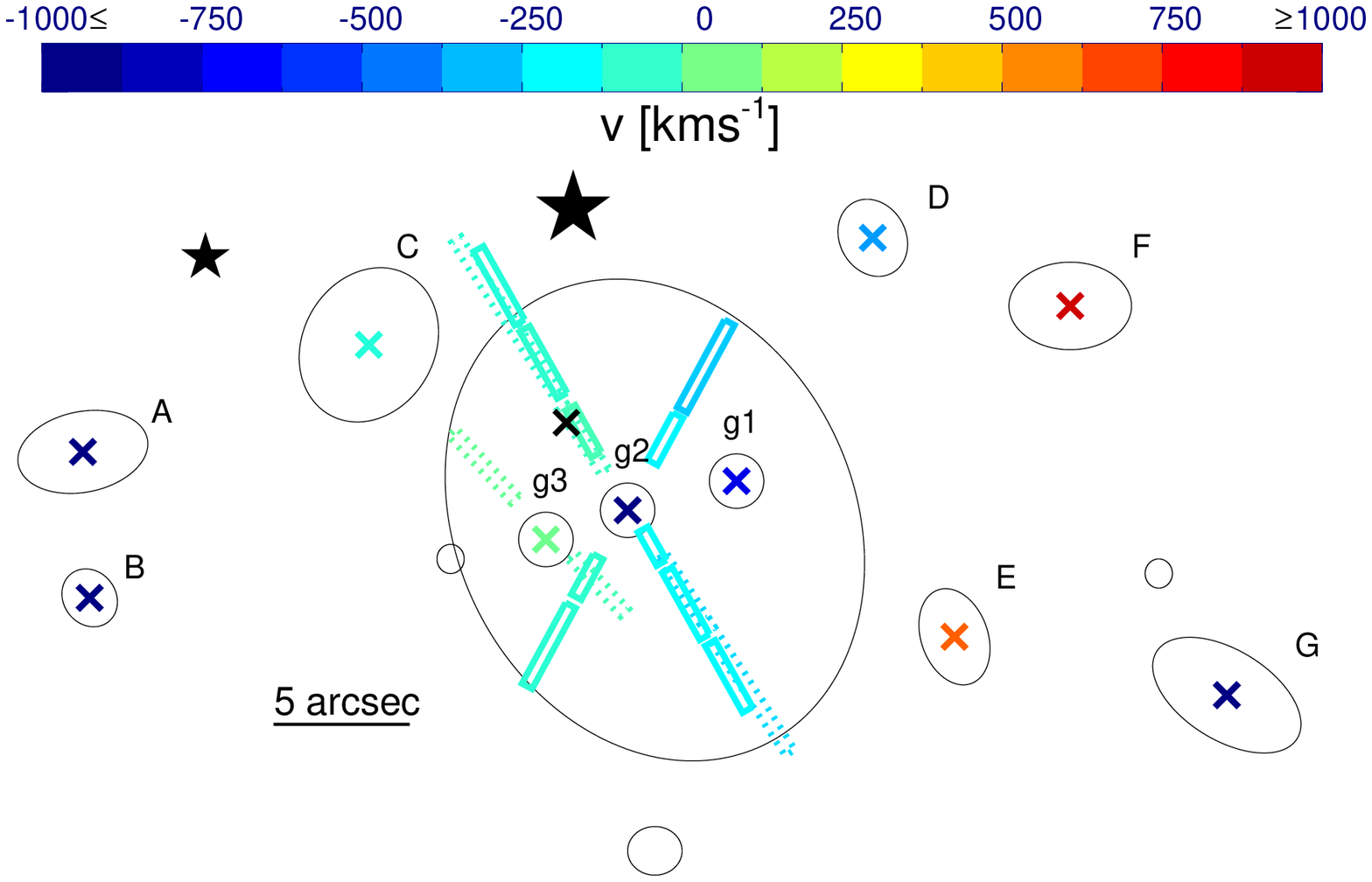}
        \includegraphics[width=0.49\textwidth]{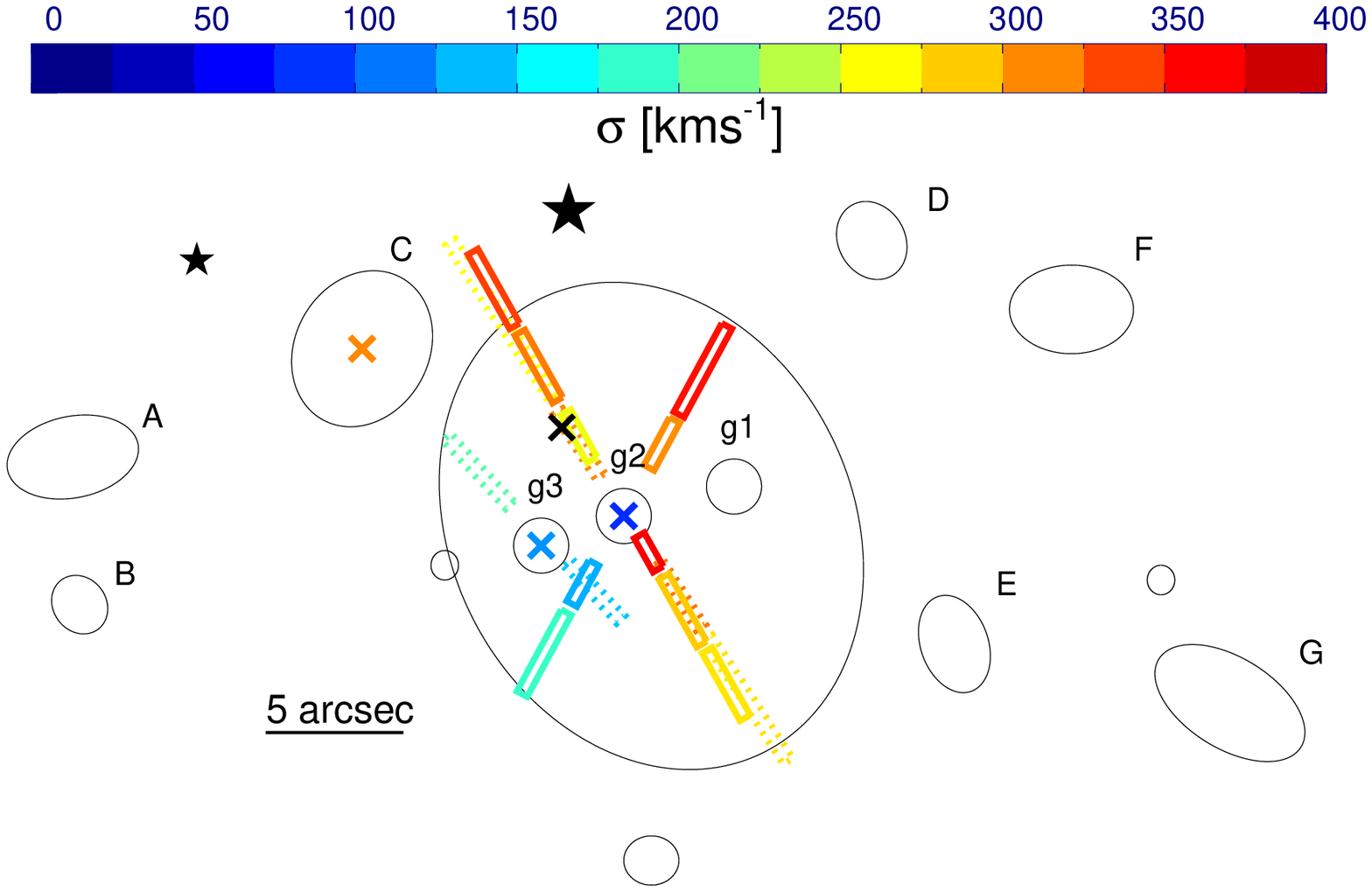}
        \caption{2-D kinematics representation at the center of Abell
          545. Ellipses indicate the positions for member galaxies and
          the star pile. Crosses indicate the measurements for member
          galaxies from the MOS campaigns, while the rectangles
          indicate the measurements from the GMOS (solid line) and
          FORS2 (dotted line) long-slits. Star symbols indicate the
          position of two spectroscopically confirmed Galactic
          stars. {\bf Top panel}: peculiar velocities at the center of Abell
          545. {\bf Bottom panel}: velocity dispersion measurements. In both
          panels, the black X symbol indicates the central position of
          the X-ray emission given by \citet{david99}. Figure scale is
          the same as in Fig. \ref{fig:longslits}. North is up, East
          to the left.}
      \label{fig:center}
    \end{figure}

    The tumultuous picture given by the velocities is reinforced by the
    velocity dispersion measurements. GMOS and FORS2 velocity
    dispersion measurements can be seen in Table \ref{table:vdps} and
    Fig. \ref{fig:center}. The GMOS major axis shows an increase in the
    velocity dispersion in the NE direction, but a decrease in the SW
    direction. The FORS2 major axis has a good agreement in the SW
    velocity dispersion and a worse agreement in the NE part,
    preferring lower velocity dispersions. The intermediate axis long
    slits show that the vicinity of the galaxy \textit{g3} has velocity
    dispersions below 200 kms$^{-1}$, and that the NW part of the star pile
    has an increasing velocity dispersion profile, similar to the NE
    part.

    The kinematic information from the long slits shows that a
    significant part of the star pile, $\sim$10 kpc in radius from
    \emph{g3}, seems to be kinematically connected to this galaxy,
    with consistent velocities and velocity dispersions. The rest of
    the star pile has very different velocities from the cluster mean
    velocity (that is the same as \textit{g3}) and larger velocity
    dispersions. This makes a dynamical connection for these parts of
    the star pile to the galaxy \textit{g3} or any surrounding galaxy
    doubtful, although it cannot be discarded that they were once part
    of \emph{g3}, but now have gained velocity due to some complex
    dynamic interactions.

   \begin{table*}
     \caption{\small{Star pile velocity dispersion measurements
         from FORS2 and GMOS longslits. Column 1 indicates the
         instrument/grism combination. Column 2 indicates the radial
         range in which the spatial binning has been done, with
         respect to the position of the nucleus the long slit
         crosses. Column 3 is the position angle of the slit, measured
         East from North (see Fig. \ref{fig:longslits} for the exact
         slit positioning). Column 4 indicates the signal-to-noise ratio of each spectra at the central wavelength of the range used for the pPXF fit. Column 5 is the velocity difference with
         the system velocity of the cluster. Column 6 indicates the
         velocity dispersion and its error.}}
     \label{table:vdps} 
\centering
\begin{tabular}{cccccc}     
  \hline \hline
  \multicolumn{1}{c}{Instrument/Grism}&
  \multicolumn{1}{c}{Radial range [kpc]}& \multicolumn{1}{c}{PA
    [degress]}& \multicolumn{1}{c}{S/N}& \multicolumn{1}{c}{$\Delta \varv$ [km s$^{-1}$]}&
  \multicolumn{1}{c}{$\sigma$[km s$^{-1}$]}\\
  \hline
  FORS2/1200R & 5--12 & 33.2  & 13&-111 $\pm$ 40 &299 $\pm$ 55\\
  Major axis  & 12--31& 33.2  & 15&-141 $\pm$ 37 &255 $\pm$ 47\\
  & 6--12 & 213.2 & 18& -312 $\pm$ 40&308 $\pm$ 35\\
  & 12--29& 213.2 & 14& -308 $\pm$ 42&265 $\pm$ 32\\
  \hline
  FORS2/1200R & 5.5-15& 43.0 &  13&-27 $\pm$ 23 &188 $\pm$ 31\\
  Interm. axis& 2--7 & 223.0 & 16 &-90 $\pm$ 39 &129 $\pm$ 20\\
  \hline
  GMOS/600B   & 5.6--10.4  & 29.3  &  9 & -91 $\pm$ 56 & 248 $\pm$ 26\\
  Major axis  & 10.4--17.6 & 29.3  & 10 & -135 $\pm$ 57 & 307 $\pm$ 34\\
  & 17.6--26.4 & 29.3  & 8& -161 $\pm$ 57 & 330 $\pm$ 44\\
  & 5.6--8.0   & 209.3 & 12& -212 $\pm$ 41 & 358 $\pm$ 45\\
  & 8.0--15.2  & 209.3 & 15& -223 $\pm$ 31 & 276 $\pm$ 29\\
  & 15.2--22.4 & 209.3 & 11& -243 $\pm$ 134 & 264 $\pm$ 36\\
  \hline
  GMOS/600B   & 5.6--9.6  & 151.6 & 11& -144 $\pm$ 55 & 120 $\pm$ 21\\
  Interm. Axis& 9.6--17.6 & 151.6 & 10& -133 $\pm$ 27 & 176 $\pm$ 15\\
  & 4.0--8.8  & 331.6 & 14& -252 $\pm$ 88 & 298 $\pm$ 21\\
  & 8.8--20.8 & 331.6 & 12& -343 $\pm$ 134 & 349 $\pm$ 25\\
  \hline
\end{tabular}
   \end{table*}

   \subsection{The star pile ``nuclei''}
   \label{sec:nuclei}
   As already mentioned in Section \ref{sec:intro} and as can be seen
   in Fig. \ref{fig:longslits}, the star pile appears to have at least
   three nuclei. Multiple nuclei are not an uncommon phenomenon in
   brightest cluster galaxies \citep[BCGS,][]{hoessel85,laine03},
   although their physical connection with the central galaxy is
   doubtful \citep{tonry85,blakeslee92}. In Fig. \ref{fig:longslits}
   these are labeled as \textit{g1--g6}, for consistency with the
   names adopted in \citetalias{salinas07}. In this section we
   describe the observations of the three brightest,
   \textit{g1-g3}. One of the aims of our MOS campaigns was to
   establish if these sources that seem embedded in the star pile
   were, in the first place, real cluster members, and secondly, if
   from their velocity it could be argued that they are actually
   nuclei of the central feature, or galaxies in the process of
   merging with the star pile.

   \citetalias{salinas07} presented photometry of the nuclei using the
   FORS1 images. The most important results are: a) the apparent
   magnitudes and colors of the nuclei are consistent with
   low-luminosity ellipticals, and b) while sources \emph{g2} and
   \emph{g3} have normal sizes, source \emph{g1} has a compact nature,
   with an effective radius of about 400 pc. Here we expand the
   discussion on the star pile nuclei using the new kinematic
   information and in the context of the MEGACAM photometry.

 \begin{table*}
   \caption{Photometry of the star pile ``nuclei'' and
     neighbor member galaxies depicted in Fig. \ref{fig:center}. 
     Column 1 is the galaxy ID. Columns 2 and 3 indicate the celestial coordinates. 
     Column 4 is the projected distance to the
     cluster center in kpcs, which is taken at the position of
     galaxy \textit{g3}. Columns 5, 6 and 7 are the $u^*g'r'i'z'$
     AB photometry. Column 8 is the recessional heliocentric velocity
     and Column 9 is the peculiar rest frame velocity.}
     \label{table:nuclei}          
\centering     
\begin{tabular}{lcccccccccr}
\hline\hline
\multicolumn{1}{c}{ID}   & \multicolumn{1}{c}{RA}&
\multicolumn{1}{c}{Dec}  & \multicolumn{1}{c}{D [kpc]}&
\multicolumn{1}{c}{$u^*$}& \multicolumn{1}{c}{$g'$}& 
\multicolumn{1}{c}{$r'$} & \multicolumn{1}{c}{$i'$}& 
\multicolumn{1}{c}{$z'$} & \multicolumn{1}{c}{$cz$ [km s$^{-1}$]}&
\multicolumn{1}{c}{$\varv_{\mathrm{pec}}$ [km s$^{-1}$]}\\ \hline
     \textit{g1}  &05:32:24.93&-11:32:38.20& 19.5&23.61&21.67 &20.53 &19.99 &19.52&46\,608$\pm$48&-792$\pm$97\\
     \textit{g2}  &05:32:25.19&-11:32:39.23& 8.2 &24.53&22.64 &21.59 &21.09 &20.66&45\,772$\pm$28&-1514$\pm$85\\
     \textit{g3}  &05:32:25.39&-11:32:40.07& 0.0 &24.97&23.07 &21.89 &21.39 &20.89&47\,511$\pm$35&-13$\pm$91\\
     \hline
     A    &05:32:26.46&-11:32:37.92&43.5 &21.22&19.08&18.11&17.61&17.59&46\,096$\pm$24&-1234$\pm$87\\
     B    &05:32:26:44&-11:32:42.39&42.4 &23.47&21.60&20.57&20.03&19.92&45\,018$\pm$28&-2164$\pm$88\\
     C    &05:32:25.79&-11:32:33.72&23.7 &20.39&18.57&17.53&16.99&16.68&47\,338$\pm$62&-162$\pm$104\\
     D    &05:32:24.63&-11:32:29.96&41.5 &21.31&19.61&18.58&18.06&17.68&47\,019$\pm$42&-437$\pm$94\\
     E    &05:32:24.44&-11:32:43.51&39.6 &21.04&19.24&18.30&17.68&17.48&48\,208$\pm$50&588$\pm$98\\
     F    &05:32:24:18&-11:32:32.16&53.6 &21.24&19.46&18.40&17.84&17.47&48\,525$\pm$42&862$\pm$94\\
     G    &05:32:23.81&-11:32:45.46&65.9 &20.53&18.44&17.35&16.79&16.50& 46\,318$\pm$35&-1042$\pm$91\\
\hline
\end{tabular}
\end{table*}

   \subsubsection{Nuclei kinematics}
   \label{sec:nuclei_kin}
 
   The radial velocity of the source \emph{g2} was already determined
   in \citetalias{salinas07} and found to be $\sim$1300 km s$^{-1}$\,
   less than the star pile velocity. This velocity has been refined
   with the new spectroscopy and now has the value of 45772$\pm$28 km
   s$^{-1}$, in very good agreement with the previous measurement. The
   very large velocity offset rules out the idea of this source as a
   nucleus of the star pile, but is not enough to classify this galaxy
   as a non-member of the cluster. Its projected position, so close to
   the galaxy cluster center, and its large line-of-sight velocity,
   provide evidence for a very radial orbit close to the line of sight
   that has reached or is about to reach the bottom of the potential
   well. Sources \emph{g1} and \emph{g3} have been also confirmed to
   be cluster members with the new measurements. Their radial
   velocities are 46\,608$\pm$48 km s$^{-1}$\, and 47\,511$\pm$35 km
   s$^{-1}$, respectively. Their peculiar velocities can be seen in
   Table \ref{table:nuclei}. The velocity of \emph{g1} also argues
   against it being a nucleus of the star pile, but the velocity of
   \textit{g3} places it exactly at the center of the velocity
   distribution, making it the most probable ``central galaxy'' of the
   cluster.

   The large exposure times for \textit{g2} and \textit{g3}, which
   were inside the slits for measuring the kinematics of the star pile
   as seen in Fig. \ref{fig:longslits}, prompted us to obtain a
   velocity dispersion for these galaxies. Unfortunately, even though
   the $S/N$ seemed sufficient, the resolution of the 1200R grism was
   not, and even though pPXF gives a good fit, it failed to give a
   meaningful value for the velocity dispersion in the case of
   \textit{g2}. Looking at the spectrum of \textit{g2}, the Mg I
   doublet (5167 and 5172\AA) looks clearly resolved, so we estimate
   the velocity dispersion for this galaxy to be $\lesssim$ 70 km
   s$^{-1}$, again consistent with a low luminosity elliptical galaxy.
   In the case of \textit{g3}, using pPXF we calculate a velocity
   dispersion of 110 $\pm$ 11 km s$^{-1}$. The pPXF fit to this galaxy
   can be seen in Fig. \ref{fig:ppxf}.
   
   The fact that the dimmer galaxy \emph{g3} has a larger velocity
   dispersion than its brighter neighbor \emph{g2} clashes with the
   expectations from scaling relations \citep{faber76}. Keeping in
   mind the difficulties in measuring total magnitudes for these
   sources we can compare the dispersion values of \emph{g2} and
   \emph{g3} with the measured velocity dispersions for nearby
   dEs. While \emph{g2} has a normal velocity dispersion for its
   brightness, \emph{g3}'s velocity dispersion is too high when
   compared with the values obtained for dEs in nearby clusters
   \citep[e.g.][]{spolaor10}. Three explanations are possible for this
   behaviour. First, that the galaxy was brighter in the past, but has
   been tidally stripped, remembering its original central velocity
   dispersion. Second, would be the expected increase in the
   dispersion induced by harrasment \citep{moore98a}, although this
   process also predicts an increase in the central luminosity density
   which would mean the galaxy had an even lower original surface
   brightness. The final possibility is that the particular position
   of the galaxy near to the cluster center makes it dark matter
   dominated even to its inner regions. This last option will be
   explored in a forthcoming paper.

   \citet{aguilar86} analyzed the effect on the density profiles of
   galaxies that have undergone a tidal encounter. One of their main
   conclusions is that strong encounters would increase the surface
   brightness and reduce the effective radius in the affected
   galaxy. The exact opposite effects are expected for weak
   encounters. The difference between strong and weak encounters lies
   on a combination of the mass ratio of the participant galaxies,
   their relative velocities and impact parameter. In the center of
   Abell 545 we can imagine a similar situation where the nuclei have
   an encounter with the cluster's dark matter halo. For the nuclei,
   their mass ratio compared to the dark halo and the impact parameter
   can be considered as equal. The main diffrence would be their
   velocities; while the high-speed galaxy \textit{g2} has been
   unaffected in its structure, the mid-speed galaxy \textit{g1} has
   presumably suffer a strong encounter, loosing its outer regions and
   mutating into a compact source.  Finally, the low-speed galaxy
   \textit{g3} has undergone a weak collision, lowering its surface
   brightness and seeming to be on the verge of complete tidal
   disruption.

   \section{The absence of a central galaxy and the origin of the star
     pile}
  \label{sec:discussion}
   
  Many lines of observational evidence point to cD galaxies as a
  family different from giant elliptical galaxies
  \citep[e.g.][]{dressler78,kormendy89,vonderlinden07}. This has
  prompted the development of many theories and models for their
  formation
  \citep[e.g.][]{merritt84a,fabian94,dubinski98,delucia07}. However,
  theoretical studies on the \textit{lack} of cD formation in some
  cluster centers is scarce.

  How does the star pile compares to the properties of cD envelopes?
  First, although it has a large size, with $\sim$ 60 kpc on its major
  axis \citep[][S+07]{struble88}, it is still small compared to cD
  sizes which can extend to hundreds of kpc
  \citep{schombert86}. Second, as we have shown in
  Sect. \ref{sec:photometry}, the light profile is much flatter than
  that of a cD envelope, and the same time, with a total luminosity of
  $L\!\sim\!5\times10^{10}$ L$_{\odot}$, it falls short compared to
  the values for cD envelopes which have luminosities between
  $10^{11}-10^{12}$L$_{\odot}$ \citep{schombert86}. On the other hand,
  the fact that the peculiar velocity of the star pile is non-zero
  does not necessarily rule it out as a cD envelope, since cD galaxies
  can have significant velocity offsets with respect to their cluster
  velocity \citep{zabludoff90, oegerle01,pimblett06}.

   One possibility, proposed by \citetalias{salinas07}, is that the
   large velocity dispersion in the center of massive clusters
   prevents the formation of a cD. Within the inner 100 kpc the
   velocity dispersion of the cluster galaxies reaches $\sim$ 900 km
   s$^{-1}$ (Salinas et al. in preparation and see also Table
   \ref{table:nuclei}). Strong tidal forces might disrupt any attempt
   of galaxy formation. Tidal forces and stripping are definitely at
   play at the center of the cluster, and this is reflected in the
   compact nature of the galaxy \textit{g1} and the extended, and
   seemingly dissolving, galaxy \textit{g3}. The luminosity of the
   star pile is larger than the predictions for the growth of a
   central galaxy through ``cannibalism'' \citep[1--2
   $L^{\star}$,][]{merritt85a,lauer88}. If stripping is the main
   origin of the star pile it is not produced by a central galaxy
   which is absent, indicating that stripping is mainly driven by the
   underlying dark matter halo.

   Intracluster light (hereafter ICL) is the name given to all the
   stellar populations that inhabit a cluster, but that are not
   gravitationally bound to any galaxy in particular but to the
   cluster potential \citep[][and references therein]{arnaboldi10}. In
   practice, to determine if stars (or planetary nebulae, or globular
   clusters) are bound or not to a galaxy is difficult, requiring
   kinematic information rarely available. Instead, ICL is usually
   handled with a working definition, for example, \citet{zibetti05}
   uses $\mu_r\!>\!25$ as a photometric threshold. Simulations predict
   that most of the ICL component is produced during the series of
   mergers which become a BCG, disfavouring a tidal origin
   \citep{abadi06,murante07}.

   As seen in previous sections, parts of the star pile can be
   considered as an extension of the galaxy named \textit{g3}, while
   other parts, especially the North-West, have a significant velocity
   difference with the systemic velocity and the velocities of
   neighbor galaxies. So perhaps we are seeing the first diffuse ICL
   component defined dynamically, besides planetary nebulae and
   globular clusters found in galaxy clusters at lower redshifts
   \citep[e.g.][]{arnaboldi04,bergond07}. But if we go back to the
   photometric definition of \citet{zibetti05}, we find that the
   entire star pile has a higher surface brightness than the
   definition with $\mu_r\!<\!24.5$. It is also higher than the values
   predicted by simulations \citep{willman04}. Perhaps ICL as bright
   as this is formed usually in massive clusters but is captured by a
   central galaxy which masks its presence.

   One further note refers to the regular appearance that the star
   pile shows, which is at odds with irregular structures such as
   tails and arcs that have been identified as ICL in nearby clusters
   \citep[e.g.][]{trentham98,gregg98,mihos05}. These cold structures
   are expected to decay rapidly ($\lesssim$ 1 Gyr) on the central
   parts of galaxy clusters \citep{rudick09}, so this can be an
   indication that the star pile is in a more evolved dynamical stage,
   but, on the other hand, the existence of significant velocity
   differences within the star pile could be considered as evidence
   for unmixed material as in the halos of nearby central galaxies
   \citep{mcneil10,ventimiglia11b}. Perhaps its seemingly smooth
   appearance is the effect of several overlapping irregular features,
   but this speculation must await support from higher resolution
   imaging (e.g. with the HST).

   The center of Abell 3827 has some resemblance to the center of
   Abell 545 \citep{carrasco10}. Abell 3827 contains a massive cD
   galaxy in the apparent process of devouring four smaller
   galaxies. The peculiar velocities of these galaxies respect to the
   systemic velocity are $\sim$100 km s$^{-1}$ for two of them,
   $\sim$300 km s$^{-1}$ for another and $\sim$1000 km s$^{-1}$ for
   the last one. So, as in Abell 545, there are galaxies that
   presumably have been slowed down by dynamical friction while the
   other is making a first passage through the cluster center. Another
   related system is the galaxy cluster CL 0958+4702, where three
   galaxies are probably merging with the BCG producing a large plume
   of stars \citep{rines07}. This diffuse halo has a similar surface
   brightness to the star pile and a very similar luminosity
   $L_r\!\sim\!5\times10^{10}$L$_{\odot}$. A key difference is the low
   velocity dispersion of the inner galaxy population in CL 0958+4702
   with $\sigma\sim$\,250\,km s$^{-1}$. Galaxy interactions are
   favored in low velocity dispersion systems
   \cite[e.g.][]{osmond04}. Another key difference is that in CL
   0958+4702, the origin of the plume has a natural explanation as
   by-product of a major merger which would eventually become a
   BCG. This is not the case in Abell 545, where there is no evidence
   of an on-going major merger, nor the possible formation of a BCG.

   \section{Summary and conclusions}

   The ``star pile'' is an extended feature of approximately
   elliptical appearance of low surface brightness with three embedded
   ``nuclei'' in the center of the massive galaxy cluster Abell 545,
   in which a cD galaxy is otherwise absent.  We present new
   kinematical details regarding the star pile and its nuclei.  Deep
   long-slit observations reveal a velocity field which clearly
   disfavor the projected central object named \emph{g2} as the
   spatially central nucleus of the star pile due to its large
   velocity offset of $-1500$ km s$^{-1}$.  Somewhat suprisingly, the
   Eastern nucleus (\emph{g3}) is kinematically connected to at least
   parts of the star pile. The velocity field of the star pile itself
   in the regions probed by the long-slits shows moderate offsets
   around $-100$ km s$^{-1}$ in the Eastern part and reaches as high
   as $-450$ km s$^{-1}$ in the North-Western part.

   The velocity dispersion neither does exhibit a simple pattern. In
   the proximity of \emph{g3} it shows moderate values of 100--150 km
   s$^{-1}$, similar to \emph{g3}, but rises along the major axis to
   the NE to values of around 300 km s$^{-1}$, but declines or stays
   constant towards the SW at values of around 200--300 km s$^{-1}$.
   The highest values of around 350 km s$^{-1}$ are found towards the
   NW.
 
   The star pile origin is most probably connected to the tidal
   stripping of at least one galaxy that now appears as its
   nucleus. It is not possible to say if the stripping proccess was
   produced by the interaction with the other embedded sources or by a
   tidal interaction with the cluster potential, which stand as the
   two most likely origin explanations. The star pile could serve as a
   probe for the inner cluster dark matter profile, only if the
   kinematics of its nucleus \emph{g3} and its near surroundings,
   which seem at rest with respect to the cluster potential, are
   considered.

\begin{acknowledgements}
  RS acknowledges support from a CONICYT doctoral fellowship. TR
  acknowledges support from the Chilean Center for Astrophysics,
  FONDAP Nr. 15010003, and from FONDECYT project Nr. 1100620. AJR was
  supported by National Science Foundation grants AST-0808099 and
  AST-0909237.
\end{acknowledgements}

\bibliographystyle{aa}
\bibliography{starpile}
\Online

\end{document}